\documentclass[epj]{svjour}
\usepackage{epsf}

\begin{document}
\title{Cold inelastic collisions between lithium and cesium in a
  two-species magneto-optical trap}

\author{U.~Schl\"oder\thanks{\emph{Present address:} Physikalisches
    Institut, Eberhard-Karls-Universit\"at, 72076 T\"ubingen, Germany}
  \and H.~Engler \and U.~Sch\"unemann\thanks{\emph{Present address:}
    BASF AG, DWL/LP - L426, 67056 Ludwigshafen, Germany} \and R.~Grimm
  \and M.~Weidem\"uller\thanks{\email{m.weidemueller@mpi-hd.mpg.de}} }
%
%
\institute{Max-Planck-Institut f\"ur Kernphysik, 69029
  Heidelberg, Germany}
\date{submitted to Euro. Phys. J. D, special issue on ``Laser Cooling
  and Trapping'', December 1998}
\abstract{We investigate collisional properties of lithium and cesium
  which are simultaneously confined in a combined magneto-optical
  trap.  Trap-loss collisions between the two species are
  comprehensively studied.  Different inelastic collision channels are
  identified, and inter-species rate coefficients as well as cross
  sections are determined. It is found that loss rates are independent
  of the optical excitation of Li, as a consequence of the repulsive
  Li$^*$-Cs interaction.  Li and Cs loss by inelastic inter-species
  collisions can completely be attributed to processes involving
  optically excited cesium (fine-structure changing collisions and
  radiative escape). By lowering the trap depth for Li, an
  additional loss channel of Li is observed which results from
  ground-state Li-Cs collisions changing the hyperfine state of
  cesium.}
%
\maketitle
\section{Introduction}
\label{sec:intro}

Cold collisions between trapped, laser-cooled atoms have been the
subject of extensive research in the past years \cite{walk94:adv}.  In
contrast to collisions of thermal atoms, the collision process between
cold atoms is extremely sensitive to the long-range part of the
inter-atomic interaction allowing precise determination of molecular
potentials \cite{mill93:prl} and atomic lifetimes \cite{mcalex96:pra}.
In the presence of light fields, molecular excitation during the
collisional process is non-negligible, leading to phenomena such as
light-induced collisions \cite{pren88:optlett,sesk89:prl},
photoassociation \cite{mill93:prl}, optical shielding of inelastic
processes \cite{bali94:epl} and formation of cold ground-state
molecules \cite{fior98:prl}.

Investigations have almost exclusively concentrated on binary
collisions between atoms of the same species
\cite{walk94:adv,supt94:optlett}.  Light-induced cold collisions
between two different species ({\em heteronuclear} collisions)
strongly differ from single-species collisions ({\em homonuclear}
collisions). The excited-state interaction potential for two different
species is of much shorter range (van-der-Waals potential $\propto
1/R^6$ at large interatomic separations $R$) than the excited state
potential for two atoms of the same species (resonant-dipole potential
$\propto 1/R^3$). In the homonuclear case, the duration of the cold
collision is much longer than the excited-state lifetimes
\cite{gems98:prl} so that the dynamics of the collisional process
greatly depends on the atom-light interaction during the collision
process. For the heteronuclear case, even a cold collision takes less
time than the lifetimes of the excited atomic states. The collision
process is therefore essentially determined by the asymptotic states
which are initially prepared, much like classical ``hot'' collisions.

However, the low temperatures of laser-cooled atoms lead to a large
extension of the molecular wavepacket formed during the cold
collision. The wavepacket spreads over typically some fraction of an
optical wavelength which is of the same order of magnitude as the
range of the interaction potentials. A light-induced cold collision
between two different species is therefore highly quantum-mechanical
with mainly the s-wave scattering distribution determining the
cross-section, in contrast to homonuclear collisions involving excited
atoms \cite{juli91:pra}.

Only recently, simultaneous trapping of two different atomic species
has been reported \cite{sant95:pra,wind98,shaf98:prl}. In this
article, we present the first investigation on inelastic cold
collisions between lithium and cesium, i.e. the lightest and the
heaviest stable alkali.  This extreme combination opens intriguing
perspectives for future experiments related to the large difference in
mass and electron affinity of the two atomic species, e.g.,
sympathetic cooling of lithium by optically cooled cesium
\cite{engl98:applphys} and the formation of cold polar molecules with
large electric dipole moment which could be trapped electrostatically
\cite{seka95:jetp}. In our experiments, both species are
simultaneously confined in a combined magneto-optical trap. Trap loss
is studied by analyzing the decay of the trapped particle number after
interruption of the loading flux, both in presence and in absence of
the other species. By choosing appropriate trap parameters, different
trap loss processes based on inelastic collisions between lithium and
cesium are identified, and the corresponding cross sections and rate
coefficients are determined.

The specific features of inelastic cold collisions between two
different species are introduced in Sec.\ \ref{sec:collisions} with
emphasis on the peculiarities of the Li-Cs system. The combined
magneto-optical trap for simultaneous confinement of lithium and
cesium is described in Sec.\ \ref{sec:comtrap}. In Sec.\ 
\ref{sec:studies} detailed quantitative studies of trap loss through
inelastic Li-Cs collisions are presented. Sec.\ \ref{sec:concl}
summarizes the main results.

\section{Two-species cold collisions}
\label{sec:collisions}

\subsection{Quasi-molecular potentials}

\begin{figure}
\epsfxsize=4cm
\vspace{5mm}
\centerline{\epsffile{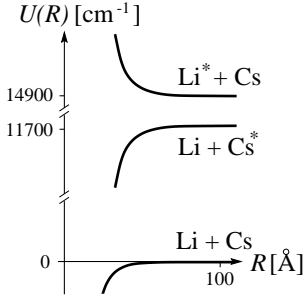}}
\caption{Long-range interaction energies plotted schematically as a
  function of internuclear distance for the ground and first excited
  states of lithium and cesium.}
\label{fig:potentials}
\end{figure}

When two cold atoms approach each other, the interaction between the
atoms leads to the formation of quasi-molecular states.  The leading
term in the long-range part of the interaction arises from the
dipole-dipole interaction.  If both atoms are in their ground state,
the potential energy is given by the well-known van-der-Waals
expression $W_{gg} = C_6 / R^6$. The coefficient $C_6$ can be
estimated by treating the two atoms A and B as simple two-level
systems with transition frequencies $\omega_i = 2 \pi c / \lambda_i$
and electric dipole moments $d_i$ ($i$ = A or B).
Second-order perturbation theory yields
\begin{equation}
C_6 \simeq - \frac{4 d_A^2 d_B^2}{\hbar(\omega_A+\omega_B)}.
\end{equation}
The van-der-Waals interaction between two ground state atoms is thus
always attractive as shown in Fig.\ \ref{fig:potentials}.

If one collision partner arrives in an excited state, the nature of
the interaction depends on whether both atoms belong to the same
species, or whether two different species collide. For the homonuclear
quasi-molecule, the interaction potential is given by the
resonant-dipole interaction $W_{ge} = C^*_3 / R^3$ with the
perturbative two-level result $C^*_3 \simeq \pm 2 d^2$. In the
heteronuclear case with atom A in the excited state and atom B in the
ground state, one obtains a van-der-Waals potential $W_{ge} = C^*_6 /
R^6$ with
\begin{equation}
C^*_6 \simeq \frac{4 d_A^2 d_B^2}{\hbar(\omega_A-\omega_B)}.
\end{equation}
The relative value of the transition energies determines the character
of the interaction. If the collision partner with the larger (smaller)
resonance frequency is excited, the interaction is generally repulsive
(attractive) as indicated in Fig.\ \ref{fig:potentials} for the case
of lithium and cesium. As we will see, this general feature of the
excited state van-der-Waals interaction has important implications on
the collisional properties of two different species.

The van-der-Waals coefficients $C_6$ and $C^*_6$ differ by the factor
$(\omega_A - \omega_B)/(\omega_A + \omega_B) \ll 1$ resulting in a
much steeper potential for the excited state than for the ground state
(see Fig.\ \ref{fig:potentials}). Although the two-level approximation
is an oversimplified model for the complex level schemes of real
atoms, the numerical values for the coefficients $C_6$, $C^*_3$ and
$C^*_6$ derived from the two-level approach reproduce the right orders
of magnitude for alkali dimers. For more accurate determination of the
long-range molecular potentials, elaborate models including spin-orbit
effects and interacting molecular states have been developed
\cite{buss87:chemphys}.

\subsection{Inelastic processes}
\label{subsec:inelcol}

Inelastic collisions in a trap lead to loss of atoms when the kinetic
energy gain of the colliding atom is larger than the trap depth. If
the energy gain is smaller than the trap depth, the atom is retained
in the trap, but the inelastic collision represents a significant
heating mechanism.  Due to the low temperatures achieved in a
magneto-optical trap (MOT), the
initial kinetic energy of the collision partners can be neglected with
respect to the interaction energy.  In the presence of light fields,
two basic processes were identified for cold inelastic collisions
involving optical excitation of the colliding pair
\cite{juli91:pra,gall89:prl}: fine-structure changing collisions (FC)
and radiative escape (RE). For two different species, an exoergic
energy-exchange reaction ${\rm A}^* + {\rm B} \rightarrow {\rm A} +
{\rm B}^* + \hbar(\omega_{\rm A} - \omega_{\rm B})$ may also take
place.  Due to the repulsive A$^*$-B potential and the large energy
defect associated with this reaction as compared to other inelastic
processes, we conjecture that this process has negligible influence.

When a photon from the light field is absorbed during the collision,
the colliding partners are accelerated towards each other on the
strongly attractive potential of the excited state.  The FC mechanism
is based on coupling of the excited molecular state to another
fine-structure or hyperfine-structure state with lower asymptotic
energy, which occurs at typical distances smaller 10\,\AA.  The
kinetic energy gain of the atom pair is the difference between the
absorbed photon energy and the energy of the lower excited
fine-structure state. The RE mechanism relies on the spontaneous
emission of a photon during acceleration on the excited molecular
potential. The gain of kinetic energy is then given by the difference
in energy between the absorbed and the emitted photon.

Both mechanisms involve one collision partner in the excited state.
The excitation probability of the collisional quasi-molecule is
largest when the detuning $\delta$ of the light field from the atomic
resonance is compensated by the interaction energy. The corresponding
internuclear distance is called the Condon point $R_C$ defined by the
condition $W_{ge}(R_C) - W_{gg}(R_C) = \hbar \delta$. For typical
detunings of a MOT ($\delta = -(1 - 6)\,\Gamma$, with $\Gamma$
denoting the inverse lifetime of the excited state), the Condon point
has values around $500 - 2000$\,\AA\ for homonuclear collisions with
the long-range $1/R^3$ reso\-nant-dipole potential, and much smaller
values around $50 - 150$\,\AA\ for heteronuclear collisions with the
shorter-range $1/R^6$ excited state van-der-Waals potential. At
distances smaller than the Condon point, the colliding atoms quickly
decouple from the light field due to the increasing energy shifts
induced by the interatomic interaction. Taking typical relative
velocities $\bar{v} = 0.1 - 1$\,m/s in a MOT and typical radiative
lifetimes $\Gamma^{-1} \simeq 30$\,ns, the semiclassical probability
of reaching small internuclear distances on an excited state molecular
potential (``survival probability'') is small for homonuclear
collisions, but might get close to unity for heteronuclear ones.

In addition to the excited-state inelastic collisions, collisions
involving both colliding atoms in the ground state may occur. In
particular, hyperfine-changing collisions (HFC) releasing the ground
state hyperfine energy, similar to the FC mechanism in the excited
state, can play a role for losses in shallow traps.

\subsection{Cold lithium-cesium collisions}
\label{subsec:licscol}

\begin{figure}
\epsfxsize=9cm 
\vspace{5mm}
\centerline{\epsffile{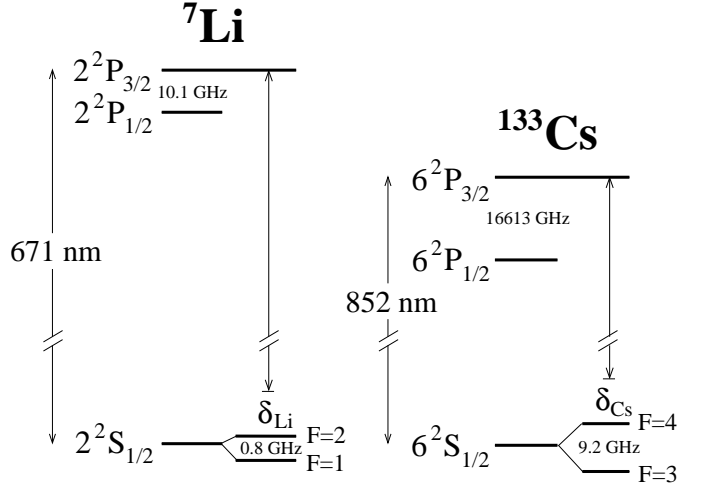}} \vspace{5mm}
\caption{Level schemes for the ground and first excited states for
  lithium and cesium.}
\label{fig:levels}
\end{figure}

The special case of a cold Li-Cs collision shows some peculiar
features. The lithium and cesium level schemes for the ground and
first excited states are shown in Fig.\ \ref{fig:levels}. Lifetimes
for the Li and Cs excited states are $(\Gamma_{\rm Li})^{-1} = 27$\,ns
and $(\Gamma_{\rm Cs})^{-1} = 30$\,ns, respectively. The $S_{1/2}$ -
$P_{3/2}$ transitions ({\it D2} line) at 671\,nm for Li and
852\,nm for Cs are used for cooling and trapping. Due to the
difference in transition energy, the quasi-molecular potential for a
Li-Cs collision is repulsive for {\em all} substates with Li$^*$+Cs
asymptotes, and attractive for {\em all} substates with Li+Cs$^*$
asymptotes \cite{buss87:chemphys} as indicated in Fig.\ 
\ref{fig:potentials}. The repulsive long-range interaction for the
Li$^*$-Cs pair has already experimentally been demonstrated by
spectroscopic measurements in a hot Li-Cs vapor \cite{vadl83:prl}.
Due to the small initial kinetic energy with respect to the
interatomic interaction potential, a Li$^*$-Cs pair is prevented from
reaching small internuclear separations where inelastic processes can
occur.

Possible inelastic channels for Li-Cs collisions are therefore
hyperfine-changing collisions for ground state Li and Cs, as well as
FC, HFC and RE collisions involving Cs$^*$. Momentum and energy
conservation require that only 5\%, i.e.\ $m_{\rm Li}/(m_{\rm Li} +
m_{\rm Cs})$ ($m_i$ = mass of the atoms), of the released energy is
transferred to the heavier collision partner Cs. When the
two-species MOT is optimized for maximum capture velocity for each
species (trap depth $\sim h\times 15\,$GHz) with $h$ = Planck's
constant), Li loss is induced by Cs when the total energy release is
larger than $\sim h \times 16$\,GHz, while the energy release for
Li-induced Cs loss has to be larger than $\sim h \times 300$\,GHz. The
hyperfine splittings of Li, Li$^*$, Cs and Cs$^*$ are therefore too
small to induce trap loss. However, for a slightly shallower Li-MOT,
the hyperfine energy of the ground-state Cs may just be sufficient to
induce loss of Li, while the Cs collisions partner remains in the MOT.

With laser cooling, much lower temperatures are achieved for Cs
($\sim 50\,\mu$K) than for Li ($\sim 1\,$mK), mainly because of
the great difference in photon recoil energy $\hbar^2 k^2/m$ with $m$
denoting the mass of the atom and $\hbar k$ the momentum of an
absorbed photon. The mean speed for Li atoms at $T_{\rm Li} =
1$\,mK is $\bar{v}_{\rm Li} = 1.7$\,m/s. This has to be compared to
the mean Cs speed $\bar{v}_{\rm Cs} = 0.1$\,m/s for a Cs
temperature of $T_{\rm Cs} = 50\,\mu$K. The Cs atoms can therefore
be considered at rest before the collision, and the mean relative
velocity $\bar{v}_{\rm LiCs}$ between cold Li and Cs is
solely determined by the Li temperature.

\section{Combined cesium-lithium trap}
\label{sec:comtrap}

\begin{figure*}
\epsfxsize=10cm
\vspace{5mm}
\centerline{\epsffile{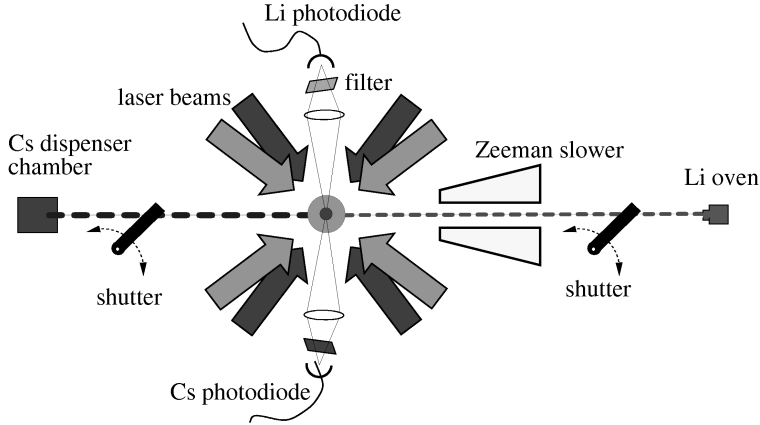}}
\caption{Experimental setup of the two-species MOT. Not shown are the
  laser beam for Li deceleration, and the two CCD cameras imaging the
  trapped atoms from different directions.}
\label{fig:setup}
\end{figure*}

A schematic view of the experiment is presented in Fig.\ 
\ref{fig:setup}. The apparatus is an extension of a MOT for Li
which is described in detail in Ref.\ \cite{schun98a:optcomm}. The
combined magneto-optical trap for lithium {\em and} cesium consists of
three mutually orthogonal pairs of counter-propagating laser beams for
each species with opposite circular polarization, intersecting at the
center of an axially symmetric magnetic quadrupole field. Field
gradients are 14\,G/cm along the vertical axis, and 7\,G/cm in the
horizontal directions. The light field of the MOT is formed by
retroreflected beams with a 1/$e^2$-diameter of 15\,mm. Total laser
power is about 15\,mW for the Cs-MOT at 852\,nm and 27\,mW for the
Li-MOT at 671\,nm. Completely separated optics is used for the two
wavelengths. The same windows are used for each trapping laser beam at
852\,nm and 671\,nm. The light is coupled into the vacuum chamber with
a small angle between the 671\,nm-beam and the 852\,nm-beam.

The laser beams are provided exclusively by diode lasers. For the
trapping of cesium, a diode laser is operated close to the
$6S_{1/2}(F=4) \rightarrow 6P_{3/2}(F=5)$ cycling transition of the
cesium {\it D2} line at 852\,nm. To avoid optical pumping into the
other hyperfine ground state, a second laser beam from a diode laser
resonant with the $6S_{1/2}(F=3) \rightarrow 6P_{3/2}(F=4)$ transition
is superimposed with the trapping beam. Both lasers are
frequency-stabilized relative to absorption lines from Cs vapor
cells at room temperature. The error signal of the servo loops is
provided by the frequency-dependent circular dichroism of Cs vapor
in a glass cell, to which a longitudinal magnetic field of some tens
of Gauss is applied. The dichroism is measured as the difference in
absorption between the left- and right-circular components of a
linearly polarized beam. Trapping of lithium is accomplished with
diode lasers in a master-slave injection-locking scheme as described
in \cite{schun98a:optcomm}. The lasers operate close to the
$2S_{1/2}(F=2) \rightarrow 2P_{3/2}$ transition and the $2S_{1/2}(F=1)
\rightarrow 2P_{3/2}$ transition, respectively, of the lithium {\it
  D2} line at 671\,nm \footnote{The excited-state hyperfine splitting
  of Li is of the same order as the natural linewidth and can thus not
  be resolved.}. One of the lasers is locked to Doppler-free
absorption lines measured by radio-frequency spectroscopy
\cite{bjor83:applphys}.  The second laser is stabilized with respect
to the first by a tunable offset-frequency lock
\cite{schun98b:revsci}.

Both MOTs are loaded from effusive atomic beams which can be
interrupted by mechanical shutters (see Fig.\ \ref{fig:setup}). The
Cs oven at a temperature of typically 85\,$^\circ$C is
continuously filled during operation by running a current of $\sim
2$\,A through a set of nine Cs dispensers. The Cs MOT
accumulates atoms from the slow velocity tail ($v \leq 10$\,m/s) of
the Maxwell distribution. Typically, close to $10^6$ atoms (at a
detuning $\delta_{\rm Cs} = - 1.5\,\Gamma_{\rm Cs}$) are trapped with
a loading time constant of several seconds. Lithium has to be
evaporated at much higher temperatures. The small mass of Li results
in much higher atom velocities. Atoms with velocity $v \leq 600$\,m/s
are decelerated in a compact Zeeman slower by an additional laser beam
at 671\,nm \cite{schun98a:optcomm}. The trapped atoms are shielded
from the Li atomic beam by a small beam block \cite{schun98a:optcomm}.
At a Li oven temperature of 450\,$^\circ$C, the loading rate is
around $10^8$\,atoms/s, yielding up to $10^9$ trapped Li atoms.
The steady-state number of trapped Li atoms can be adjusted over
a wide range by decreasing the loading flux through attenuation of the
Zeeman-slowing laser beam. Densities for the Cs and the Li
MOT range between $10^9$ and $10^{10}$\,atoms/cm$^3$.

The fluorescence of the trapped atoms is monitored by two calibrated
photodiodes with narrow-band interference filters at 852\,nm and
671\,nm, respectively. Shape and position of the two atomic clouds are
measured with two CCD cameras. From these measurements, the number of
trapped atoms and the density are determined. The cameras are looking
from different directions yielding 3D information on the position of
the Li and Cs cloud.

The clouds are not necessarily overlapping. To superimpose both
clouds, we found it most simple and reproducible to shift the Li
onto the Cs cloud by slightly focusing the retroreflected beams at
671\,nm and thus introducing a controlled radiation-pressure
imbalance. The Li cloud turned out to be more sensitive to a
radiation-pressure imbalance than the Cs cloud.

\begin{figure}
\epsfxsize=5cm
\vspace{5mm}
\centerline{\epsffile{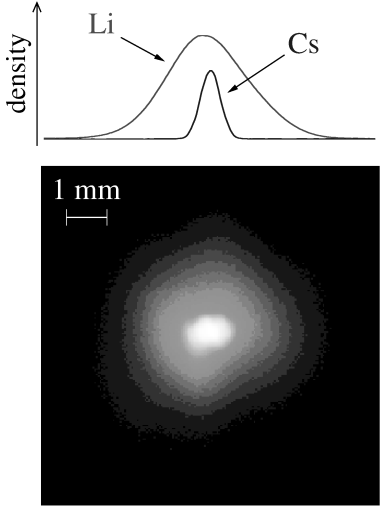}}
\caption{Camera picture of simultaneously trapped lithium and cesium
  atoms. The density distributions shown above are measured separately
  for lithium and cesium using narrow-band filters in front of the CCD
  camera.}
\label{fig:csliMOT}
\end{figure}

Cooling in the Li-MOT is based on Doppler forces
\cite{schun98a:optcomm}, while polarization-gradient forces are acting
on the trapped Cs \cite{stea92:josab}. As a consequence of the
different mechanisms, temperatures of the Li cloud in the MOT are
above the Doppler temperature ($T_{Li} \approx 1$\,mK), while the
Cs cloud is cooled to sub-Doppler temperatures ($T_{Cs} \approx
50\,\mu$K). Fig.\ \ref{fig:csliMOT} shows a fluorescence picture of
simultaneously trapped Li and Cs atoms. Due to its much lower
temperature the Cs cloud occupies a much smaller volume than the
Li cloud, as indicated by the density profiles in Fig.\ 
\ref{fig:csliMOT}. This particular property of the Li/Cs system
greatly simplifies quantitative collisional studies.

\begin{figure}
  \epsfxsize=5cm 
\vspace{5mm}
\centerline{\epsffile{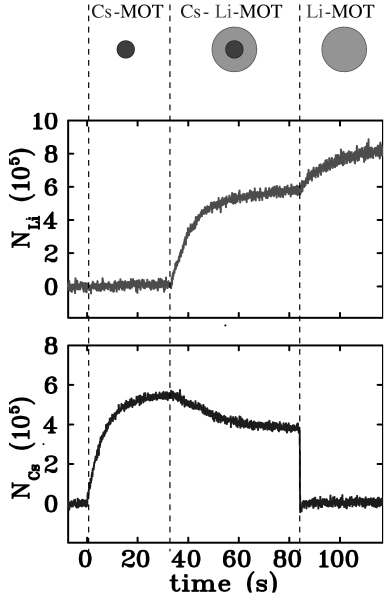}}
\caption{Temporal evolution of the number of trapped lithium and
  cesium atoms during loading of the two-species MOT with and without
  the other species present.}
\label{fig:numb_atoms}
\end{figure}

Binary inelastic collisions between lithium and cesium lead to loss
from the two-species MOT. One indication of this loss is a decrease of
the steady-state particle number of one species when the other species
is also loaded into the combined MOT. In Fig.\ \ref{fig:numb_atoms},
the temporal evolution of the trapped particle number during loading
is shown to illustrate the influence of inter-species collisions.
First, only Cs is loaded into the MOT until the steady-state number is
reached. Then, as Li is also filled into the trap by opening the
atomic beam shutter, the number of trapped Cs decreases which
indicates inelastic Li-Cs collisions resulting in a trap loss of Cs.
After a new steady state has established, loading of Cs is stopped by
shuttering the Cs beam, and the light field at 852\,nm is
interrupted for a short moment, resulting in quick escape of all Cs
atoms. Without Cs, the number of trapped Li further increases which
shows that inter-species collisions also induce Li loss.

\section{Quantitative studies}
\label{sec:studies}

\subsection{Measurement procedures}

Inelastic collisions can be studied quantitatively by measuring rate
coefficients for the loss of particles from the trap. The temporal
evolution of the trapped particle number $N_A$ for the species A under
the presence of species B is described by the rate equation
\begin{equation}
\label{eq:rateeq}
\frac{d N_{\rm A}}{dt} = L_{\rm A} - \alpha_{\rm A} N_{\rm A} -
  \beta_{\rm A} \int n_{\rm A}^2\,d^3\!r - \gamma_{\rm AB} \int
  n_{\rm A} n_{\rm B}\,d^3\!r
\end{equation}
where $n_{\rm A,B}$ denote the local densities and $L_{\rm A}$ the
loading rate for species A. The loss rate coefficient $\alpha_{\rm A}$
in the second term of Eq.\ \ref{eq:rateeq} characterizes trap loss by
collisions with background particles. Inelastic binary collisions
between trapped particles are described by the last two terms in Eq.\ 
\ref{eq:rateeq}. The rate coefficients $\beta_{\rm A}$ and
$\gamma_{\rm AB}$ denote the loss rate coefficient for collisions
between atoms of the same species and between different atomic
species, respectively. These coefficients can be expressed in terms of
trap-loss cross sections $\beta_{\rm A} = \bar{v}_{\rm AA} \sigma_{\rm
  A}$ and $\gamma_{\rm AB} = \bar{v}_{\rm AB} \sigma_{\rm AB}$ where
$\bar{v}_{\rm AA}$ and $\bar{v}_{\rm AB}$ denote the relative speed
between two atoms of species A and between species A and B,
respectively.

The rate coefficients for trap loss in Eq.\ \ref{eq:rateeq} can be
inferred from the decay of the fluorescence signal after interruption
of the loading flux for species A ($L_{\rm A} = 0$ in Eq.\ 
\ref{eq:rateeq}).  Species B is still continuously loaded into the
two-species MOT ($L_{\rm B} \neq 0$). The fluorescence signal from the
MOT is proportional to the particle number when the cloud of trapped
atoms is not optically thick which is well fulfilled in all our
measurements.  Analysis of the data is simplified by the fact that,
for low numbers of trapped particles, the cloud extension is
determined solely by the temperature ({\em temperature-limited
  regime}) \cite{town95:pra}. In this regime, the root-mean-square
radius $r_{\rm A}$ of the Gaussian spatial density distribution is
independent of the number of particles which we have carefully checked
for our two-species MOT \cite{schlo98:diplom}.  Thus, the quadratic
loss term can be written as $- \beta_{\rm A} N_{\rm A}^2 / \sqrt{8}
V_{\rm A}$ where we call $V_{\rm A}=\left(\sqrt{2 \pi} r_{\rm
    A}\right)^3$ the volume of the species A cloud.  The volume $V_A$
stays constant during the decay of the trapped particle number. In
addition, the Li cloud is generally much larger than the Cs cloud (see
Fig.\ \ref{fig:csliMOT}). The third term in Eq.\ (\ref{eq:rateeq})
therefore simplifies to $-\gamma \hat{n}_{\rm Li} N_{\rm Cs}$ where
$\hat{n}_{\rm Li} = N_{\rm Li}/V_{\rm Li}$ denotes the Li peak
density.

With these simplifications, the decay of the trapped particle number
$N_{\rm A}$ is described by
\begin{equation}
\label{eq:rateeqdecay}
\frac{d N_{\rm A}}{dt} = - \left(\alpha_{\rm A} +  \frac{\gamma_{\rm
      AB}}{V_{\rm Li}} N_{\rm B} \right) N_{\rm A} -
  \frac{\beta_{\rm A}}{\sqrt{8} V_{\rm A}} N_{\rm A}^2
\end{equation}
where A and B stand for Li or Cs. In the general case, $N_{\rm A}$ and
$N_{\rm B}$ are coupled by the inelastic inter-species collisions (see
Fig.\ \ref{fig:numb_atoms}). However, when the loading flux for
species B is large compared to the loss rate by inter-species
collisions, i.e.\ $L_{\rm B} >> \gamma_{\rm BA} N_{\rm A} N_{\rm B}/
V_{\rm Li}$, the steady-state particle number $N_{\rm B,0}$ is not
influenced by the presence of species A. In this case, the rate
equations for A and B become decoupled, and Eq.\ \ref{eq:rateeqdecay}
has the simple analytical solution
\begin{equation}
\label{eq:solution}
N_{\rm A}(t) = \frac{N_{\rm A,0} e^{-\tilde{\alpha}_{\rm A} t}}{1 +
  \frac{N_{\rm A,0}}{\sqrt{8} V_{\rm A}} \frac{\beta_{\rm
      A}}{\tilde{\alpha}_{\rm A}} \left(1 - e^{-\tilde{\alpha}_{\rm A} t}\right)}
\end{equation}
with the effective decay rate coefficient $\tilde{\alpha}_{\rm A} =
\alpha_{\rm A} + \gamma_{\rm AB} N_{\rm B,0} / V_{\rm Li}$.

\begin{figure}
\epsfxsize=7cm
\vspace{5mm}
\centerline{\epsffile{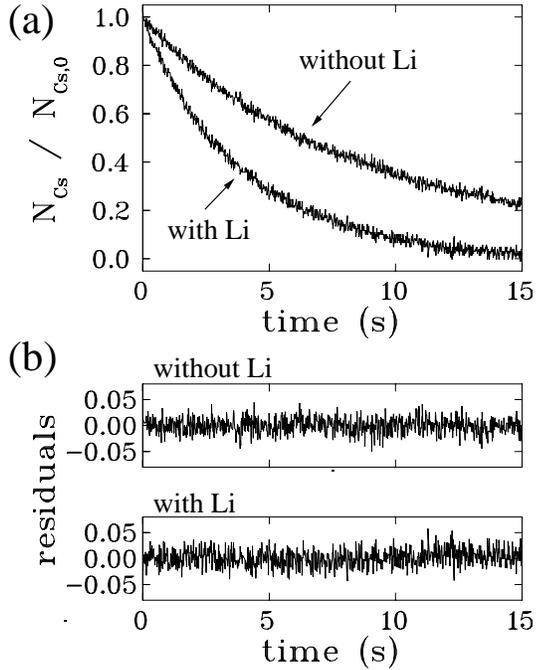}}
\caption{Decay of fluorescence for Cs in the two-species MOT with
  and without trapped Li ($\delta_{\rm Cs} =
  -1.5\,\Gamma_{\rm Cs}$, $\delta_{\rm Li} = -3\,\Gamma_{\rm Li}$).
  The fluorescence signal is proportional to the number of trapped
  atoms $N_{\rm Cs}$. The lower graphs show the residuals from a fit
  to Eq.\ \ref{eq:solution}.}
\label{fig:decay}
\end{figure}

The coefficients $\alpha_{\rm A}$ and $\beta_{\rm A}$ in
Eq.~(\ref{eq:rateeqdecay}) are determined by fitting
Eq.~(\ref{eq:solution}) to the fluorescence decay {\em without} the
species B loaded into the MOT ($N_{\rm B,0} = 0$). A typical
example is depicted in Fig.\ \ref{fig:decay}. We have found no
influence of the trapping light for species B on the rate
coefficients for species A. However, to exclude any possible
ambiguities, the trapping light for species B is not interrupted
during the measurements without species B being loaded into the
trap. In addition, we observe no influence of the species B atomic
beam on the decay characteristics of species A when opening the
beam shutter and interrupting one arm of the MOT laser beams so
that no species B atoms are trapped.

As shown in Fig.\ \ref{fig:decay}, the fluorescence decay changes
significantly when the second species is also confined in the
two-species MOT. By adjusting the loading fluxes, the particle
number $N_{\rm B,0}$ is decoupled from the decay of species A which
is checked by monitoring the fluorescence of species B. Besides the
initial particle number $N_{\rm A,0}$, the inter-species rate
coefficient $\gamma_{\rm AB}$ is the only free fitting parameter
used since the single-species parameters $\alpha_{\rm A}$ and
$\beta_{\rm
  A}$ are kept fixed to the values determined without B. In this way,
determination of $\gamma_{\rm AB}$ is uncorrelated to the
evaluation of $\alpha_{\rm A}$ and $\beta_{\rm A}$ which greatly
reduces the fitting errors.

Experimental errors in the determination of the particle numbers
$N_{\rm A,0}$ and $N_{\rm B,0}$ are 25\%, while the trap volumes
$V_{\rm A}$ and $V_{\rm B}$ are accurate to within 15\%. The errors
given in the following refer to the combination of experimental errors
with the statistical errors of the fitting procedures. Not included
are possible systematic errors in the particle number determination
which we estimate to about 50\%. Comparison of the values for
$\beta_{\rm A}$ from our measurements with previous values measured in
single-species MOTs provide a consistency check for our data analysis
and the influence of systematics.

\subsection{Lithium-induced cesium loss}
\label{subsec:lics}

Following the procedures described in the preceding section, we
have studied the trap loss of Cs atoms resulting from inelastic
Li-Cs collisions. Li and Cs are loaded for about 30\,s to their
steady-state particle numbers ($N_{\rm Cs,0} \simeq 10^6$ at
$\delta_{\rm Cs} = -1.5 \Gamma_{\rm Cs}$, $N_{\rm Li,0} \approx
10^8$ at $\delta_{\rm Li} = -3 \Gamma_{\rm Li}$). The decay of the
Cs fluorescence is monitored with and without Li after the Cs
loading flux is interrupted. In addition, the Li fluorescence is
observed during the decay of the Cs fluorescence to verify that
$N_{\rm Li,0}$ is independent of $N_{\rm Cs}$. For each set of
measurements, a camera picture is taken to measure the spatial
volume $V_{\rm Li}$ of the Li cloud.

We first investigate the influence of the population of the lithium
$2P_{3/2}$ excited state on $\gamma_{\rm CsLi}$. After the Cs
loading is interrupted, the average Li excitation is adjusted by
periodically chopping the trapping light. The chopping frequency of
$100$\,kHz is slow compared to the internal dynamics of the Li
atoms determined by $\Gamma_{\rm Li}$, but fast compared to the
dynamics of the trapped particles.  Therefore, the average excitation
of the Li atoms scales linearly with the ratio of the on/off time
intervals (duty cycle) \footnote{The Li loading rate changes with
  the duty cycle resulting in a decay of $N_{\rm Li,0}$ to a new
  steady-state value.  Therefore, Eq.\ \ref{eq:solution} can not be
  used in this measurement. The Li decay is measured by monitoring the
  Li fluorescence. The observed decay of $N_{\rm Li,0}$ is
  incorporated into Eq.\ \ref{eq:rateeqdecay}.}. At 100\,\% duty cycle
(no chopping), the average excited-state population $\Pi_{{\rm Li}^*}$
is 0.06(1) at a detuning $\delta_{\rm Li} = -3\,\Gamma_{\rm Li}$. To
determine the population, the fluorescence rate is measured as the
function of detuning for a fixed number of trapped atoms. The
excited-state population is then deduced from two-level theory by
fitting a Lorentzian to the data. The volume $V_{\rm Li}$ of the
Li MOT increases with decreasing duty cycle from 1.3(1)\,mm$^3$
at 100\,\% duty cycle to 3.0(3)\,mm$^3$ at 30\%. The Li
temperature does not change significantly with the duty cycle ($T_{\rm
  Li} = 0.9(2)$\,mK at $\delta_{\rm Li} = -3\,\Gamma_{\rm Li}$).

\begin{figure}
\epsfxsize=8cm
\vspace{5mm}
\centerline{\epsffile{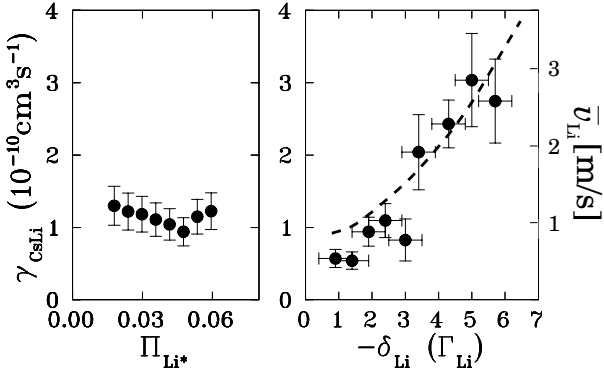}}
\caption{Left graph: Rate coefficient for lithium-induced cesium loss
  $\gamma_{\rm CsLi}$ as a function of the lithium population of the
  excited $2P_{3/2}$ state ($\delta_{\rm Cs} = -1.5\,\Gamma_{\rm Cs}$,
  $\delta_{\rm Li} = -3\,\Gamma_{\rm Li}$). The excitation was
  controlled by square-wave modulation of the lithium trapping light.
  Right graph: Rate coefficient for lithium-induced cesium loss as a
  function of the lithium detuning $\delta_{\rm Li}$ ($\delta_{\rm Cs}
  = -1.5\,\Gamma_{\rm Cs}$). The dashed line corresponding to the
  right abscissa shows the variation of the mean lithium velocity
  $\bar{v}_{\rm Li}$ with detuning.}
\label{fig:csdetunli}
\end{figure}

The data presented in the left graph of Fig.~\ref{fig:csdetunli}
show that the Cs loss rate coefficient $\gamma_{\rm CsLi}$ has the
constant value of $\gamma_{\rm CsLi} = 1.1(2) \times
10^{-10}\,$cm$^3$/s at $\delta_{\rm Cs} = -1.5 \Gamma_{\rm Cs}$ and
$\delta_{\rm Li} = -3 \Gamma_{\rm Li}$). The coefficient exhibits
no significant dependence on the Li excitation. This observation
can be regarded as a direct consequence of the repulsive
interaction between excited Li and ground-state Cs as discussed in
Sec.\ \ref{subsec:licscol}. Excited Li in the MOT therefore
does not contribute to the trap loss of Cs.

In a second set of measurements, the dependence of $\gamma_{\rm CsLi}$
on the detuning $\delta_{\rm Li}$ of the trapping light for Li is
investigated. As shown in the right graph of Fig.\ 
\ref{fig:csdetunli}, the inter-species rate coefficient steadily
increases with increasing detuning from
$0.6(1)\times10^{-10}$\,cm$^3$/s at $\delta_{\rm Li} = -1\,
\Gamma_{\rm Li}$ to $3.0(6)\times10^{-10}$\,cm$^3$/s at $\delta_{\rm
  Li} = -6\,\Gamma_{\rm Li}$.  Changing $\delta_{\rm Li}$ has two
major consequences: the temperature of Li increases with
increasing detuning as demonstrated in \cite{schun98a:optcomm}, and
the excitation probability for Li is modified. The dashed line in
Fig.\ \ref{fig:csdetunli} gives the dependence of $\bar{v}_{\rm
  Li} = (\frac{8}{3} k_B T_{\rm Li} / m_{\rm Li})^{1/2}$ on the
Li detuning as measured for our Li MOT \cite{schun98a:optcomm}.
As discussed in Sec.\ \ref{subsec:licscol}, the Li velocity
determines the average relative velocity between lithium and cesium
$\bar{v}_{\rm LiCs} \approx \bar{v}_{\rm Li}$.  Since the change in
$\bar{v}_{\rm Li}$ essentially reproduces the measured trend of the
rate coefficient, it follows that the cross section $\sigma_{\rm CsLi}
= \gamma_{\rm CsLi} / \bar{v}_{\rm CsLi}$ for Li-induced Cs loss
is independent on the Li detuning ($\sigma_{\rm CsLi} =
0.7(2)\times10^4$\,\AA$^2$ at $\delta_{\rm Cs} = -1.5\,\Gamma_{\rm
  Cs}$). The data again indicate that the Li excitation plays no role
in inelastic Li-Cs collisions.

The rate coefficient $\gamma_{\rm CsLi} =
1.1(2)\times10^{-10}$\,cm$^3$/s at $\delta_{\rm Cs} = -
1.5\,\Gamma_{\rm Cs}$ is about one order of magnitude larger than the
homonuclear coefficient $\beta_{\rm Cs} =
2.0(4)\times10^{-11}$\,cm$^3$/s measured under the same conditions but
without lithium in the trap \cite{schlo98:diplom} \footnote{The value
  of $\beta_{\rm Cs}$ is consistent with earlier measurements in a
  Cs MOT by Sesko {\em et al.} \cite{sesk89:prl}.}. The
corresponding cross sections $\sigma_{\rm CsLi} =
0.7(2)\times10^4$\,\AA$^2$ and $\sigma_{\rm Cs} =
2.0(4)\times10^4\,$\AA$^2$, however, are of the same order of
magnitude due to the much smaller relative velocities of the Cs
atoms (see Sec.\ \ref{subsec:licscol}).

\begin{figure}
\vspace{5mm}
  \epsfxsize=8cm \centerline{\epsffile{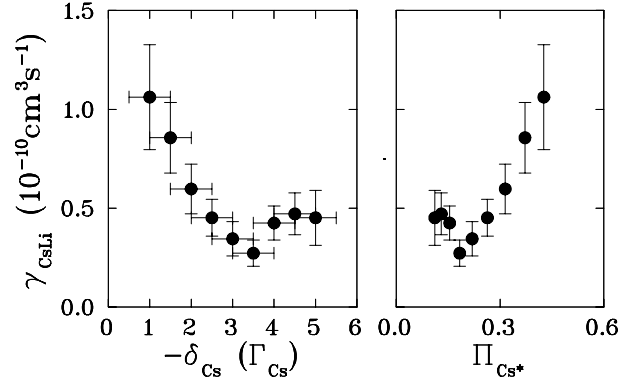}}
\caption{Left graph: Dependence of the loss rate coefficient for
  lithium-induced cesium loss $\gamma_{\rm CsLi}$ on the cesium
  detuning $\delta_{\rm Cs}$ ($\delta_{\rm Li} = -3\,\Gamma_{\rm
    Li}$). Right graph: Same data, but plotted versus the average
  population of the Cs $6P_{3/2}$ state as determined from the
  detuning.}
\label{fig:csdetuncs}
\end{figure}

To investigate the influence of optical Cs excitation, the
detuning $\delta_{\rm Cs}$ of the Cs-trapping light is switched to
a given value after interruption of the Cs loading flux. The
Li detuning is kept fixed at $\delta_{\rm Li} =
-3\,\Gamma_{\rm Li}$ resulting in a constant number of $N_{\rm Li,0}
= 5(2)\times10^6$ Li atoms at a density $\hat{n}_{\rm Li} =
1.7(7)\times10^9$\,cm$^{-3}$ in the MOT. As shown in the left graph
in Fig.\ \ref{fig:csdetuncs}, one observes a decrease of the rate
coefficient by a factor of five for increasing detuning
$\delta_{\rm
  Cs}$. At higher detuning, $\gamma_{\rm CsLi}$ rises up again.

Changing the detuning of the Cs MOT has two effects: excitation
of the Cs $6P_{3/2}$ state depends on the detuning, and the Cs trap
becomes shallower at larger detunings. In addition, the Cs
temperature becomes lower at larger detuning due to
polarization-gradient cooling \cite{stea92:josab}, but this does
not effect the rate coefficient since the relative velocity is
determined by the Li temperature only. The increase of the
rate coefficient at $-\delta_{\rm Cs} \geq 4\,\Gamma_{\rm Cs}$ can
be attributed to the decrease of the MOT depth. The MOT might
eventually become shallow enough to allow for trap loss due to
Li-Cs collisions changing the Cs hyperfine structure. From a
simplified yet realistic picture for the capture range of the
Cs MOT \cite{lind92:pra}, we expect this process to become
relevant at detunings below $\approx - 4\,\Gamma_{\rm Cs}$
consistent with the observed increase of $\gamma_{\rm CsLi}$. The
decrease of $\gamma_{\rm CsLi}$ with detuning for $-\delta_{\rm Cs}
\leq 4\,\Gamma_{\rm Cs}$, however, must be related to the change in
the Cs excitation.

The average population of the Cs in the $P_{3/2}$ state in the
MOT decreases with the detuning. As described in Sec.\
\ref{subsec:inelcol}, the relevant inelastic processes for trap
loss occur at internuclear distances around 10\,\AA. Due to the
large relative velocities of about 1\,m/s, excitation of Cs
survives over an internuclear distance of about 300\,\AA. The
Cs atoms might therefore be excited at separations before the
interatomic interaction energy becomes relevant ($R_C \approx
100$\,\AA), and still reach the inner interaction zone. The
modification of the excitation probability by the interaction
potential therefore plays a minor role for the probability for an
excited atom to reach the inner interaction zone in the excited
state. It seems therefore appropriate to expect a linear increase
of $\gamma_{\rm CsLi}$ with the average excited state population in
the MOT.

To support this picture, the right graph in Fig.\ \ref{fig:csdetuncs}
shows the same data as in (a), but now plotted versus the average
population $\Pi_{{\rm Cs}^*}$ of the Cs $P_{3/2}$ state.  The
excited-state population is measured as described above for Li.  The
rate coefficient scales proportional to the average $P_{3/2}$
population indicating that excitation relevant for the inelastic
processes indeed occurs at large internuclear distances where the
modification of the energy through the quasi-molecular potential can be
neglected. In addition, the rapid decrease of the rate coefficient
with decreasing Cs excitation shows, that inelastic Li-Cs$^*$
collisions are the main channel for Li-induced Cs trap loss.

The strong decrease of the rate coefficient with increasing detuning
constitutes an important difference to homonuclear collisions where
the trap loss rate is found to increase with increasing detuning
\cite{sesk89:prl}.  In the homonuclear case, the colliding atoms are
decoupled from the light field already at distances around 1000\,\AA\
due to the long-range resonant-dipole interaction. The rate
coefficient for homonuclear collisions can be increased by primarily
exciting the atoms at smaller internuclear separations, i.e. at larger
detunings from resonance for the attractive interaction potential,
resulting in a larger survival probability.

\subsection{Cesium-induced lithium loss}
\label{subsec:csli}

The investigation of Cs-induced Li loss from the MOT proceeds
similar to the experiments on Li-induced Cs loss. Cesium is
permanently loaded, and at a given moment the Li loading flux
is interrupted for a measurement of the Li trap decay. It has now
to be ensured that the steady-state number $N_{\rm Cs,0}$ of
trapped Cs is independent of the number of trapped Li $N_{\rm Li}$,
i.e., the Cs loading rate has to be chosen large compared to
the Li-induced Cs loss rate. By decreasing the Li loading
flux, the steady-state Li particle number in the MOT is
adjusted to values comparable to the largest achievable Cs
particle number ($N_{\rm Li,0} \approx N_{\rm Cs,0} \simeq 10^6$).
Under these conditions, the Cs fluorescence shows only a
marginal dependence on the number of trapped Li atoms so that
Eq.\ \ref{eq:solution} can be used to analyze the data. At the
corresponding low Li densities, the decay of the Li
fluorescence was found to be purely exponential indicating that
quadratic Li loss can be neglected ($\beta_{\rm Li} N_{\rm
Li,0}/\sqrt{8} V_{\rm Li} \ll
\alpha_{\rm Li,eff}$ in Eq.\ \ref{eq:solution}).

\begin{figure}
\epsfxsize=5cm
\vspace{5mm}
\centerline{\epsffile{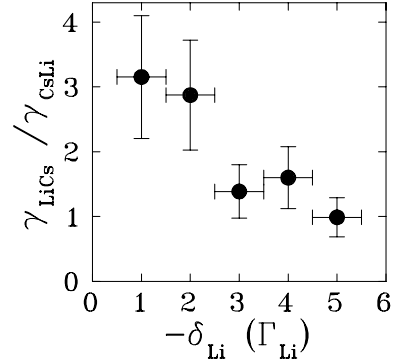}}
\caption{Ratio between the loss coefficients for cesium-induced
  lithium loss $\gamma_{\rm LiCs}$ and lithium-induced cesium loss
  $\gamma_{\rm CsLi}$ versus lithium detuning $\delta_{\rm Li}$
  ($\delta_{\rm Cs} = -1.5\,\Gamma_{\rm Cs}$).}
\label{fig:gamlicsa}
\end{figure}

From energetical considerations discussed in Sec.\
\ref{subsec:licscol}, trap escape of a Cs atom through an inelastic
Li-Cs collision has to be accompanied by the loss of the involved
Li atom, since the largest share of the released energy is taken by
the Li. In Fig.\ \ref{fig:gamlicsa}, the ratio between the
loss rate coefficient for a Cs-induced Li loss
$\gamma_{\rm LiCs}$ and the coefficient for a Li-induced
Cs loss $\gamma_{\rm CsLi}$ is depicted as function of the
Li detuning $\delta_{\rm Li}$. For $-\delta_{\rm Li} \geq 3
\Gamma_{\rm Li}$ one observes $\gamma_{\rm
  LiCs} \approx \gamma_{\rm CsLi}$, which shows that both collisions
partners simultaneously leave the trap. Since the Cs trap escape is
essentially determined by collisions involving excited Cs, this
collision channel is also the main source for Li loss.

Interestingly, at smaller detunings, an additional loss channel for
Li atoms opens which is not accompanied by the loss of the Cs
atoms. A possible process releasing sufficient energy for the escape
of Li without providing enough energy for Cs is represented
by inelastic collisions between Cs and Li both in the ground
state (hyperfine-changing collisions, see Sec.\ \ref{subsec:inelcol}).
In particular, in the MOT nearly all ground-state Cs atoms occupy
the $6S_{1/2}(F = 4)$ level.  Collisions changing the hyperfine state
of the Cs would transfer around $h\times9$\,GHz of energy to
the Li atom, which corresponds roughly to the Li trap depth
at small detunings \footnote{For Li, the trap depth steadily
  increases with detuning in the parameter ranges considered here
  \cite{schun98a:optcomm}.}.

\begin{figure}
\epsfxsize=5cm
\vspace{5mm}
\centerline{\epsffile{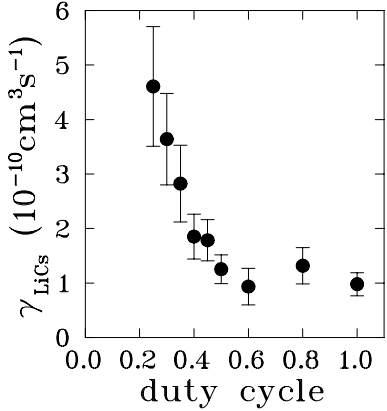}}
\caption{Loss rate coefficient of cesium-induced lithium loss
  $\gamma_{\rm LiCs}$ versus the duty cycle of square-wave
  modulation of the trap light  ($\delta_{\rm Cs} =
  -3\,\Gamma_{\rm Cs}$, $\delta_{\rm Li} = -4\,\Gamma_{\rm Li}$).}
\label{fig:gamlicsb}
\end{figure}

To further investigate the hypothesis that the additional Li loss is a
manifestation of hyperfine-changing collisions, we have changed the
Li trap depth by square-wave modulation of the Li trapping light
\cite{kawa93:pra} as explained in the preceding section. At full duty
cycle, the trap depth is estimated from the laser power to be around
15\,GHz. Lowering the duty cycle thus reduces the trap depth
sufficiently to allow for the inset of loss through
hyperfine-structure change of the Cs ground state. As shown in Fig.\ 
\ref{fig:gamlicsb} for $\delta_{\rm Li} = -4\,\Gamma_{\rm Li}$, the
loss rate coefficient for Cs-induced Li loss $\gamma_{\rm
  LiCs}$ drastically increases when the duty cycle is reduced below a
critical value of $\approx 40\%$. At a duty cycle of 20\%, a rate
coefficient of $\gamma_{\rm LiCs} = 5(1)\times10^{-10}$\,cm$^3$/s is
measured, corresponding to a cross section of $\sigma_{\rm LiCs} =
3(1)\times10^4$\,\AA$^2$.

The square-wave modulation method has formerly been used to identify
fine-structure changing collisions in a pure Li MOT which releases
5\,GHz energy to each Li collision partner \cite{kawa93:pra}. In
these experiments, a sudden increase of the rate coefficient
$\beta_{\rm Li}$ with decreasing duty cycle was observed when the duty
cycle was lowered beyond the value corresponding to 5\,GHz trap depth.
We have performed equivalent measurements on $\beta_{\rm Li}$ for the
same trap parameters as the data set shown in Fig.\ 
\ref{fig:gamlicsb}, but with maximum Li loading flux to achieve large
numbers of trapped Li ($N_{\rm Li,0}\approx 10^8$). This leads to a
measurable influence of $\beta_{\rm Li}$ on the trap loss
\cite{schlo98:diplom}. The rate coefficient $\beta_{\rm Li}$ increases
from $5(2)\times10^{-12}$\,cm$^3$/s for duty cycles between 60\% and
100\% to $\sim 1\times10^{-10}$\,cm$^3$/s at duty cycles below 40\%
\footnote{These values are consistent with rate coefficients from
  Li-MOT measurements by Kawanake {\em et al.}  \cite{kawa93:pra} and
  Ritchie {\it et al.}  \cite{ritch95:pra}.}. We find that the sudden
increase in $\beta_{\rm Li}$ sets in at a slightly lower critical duty
cycle than the increase of $\gamma_{\rm LiCs}$ shown in Fig.\ 
\ref{fig:gamlicsb}.  This indicates, that the corresponding kinetic
energy gain transferred to the lithium through an inelastic
ground-state Li-Cs collisions must be larger than $h\times5$\,GHz. The
only process releasing sufficient energy to explain the observations
is therefore an inelastic collision changing the Cs hyperfine
state.  Note, that inelastic Li$^*$-Cs collisions
changing the Li excited-state fine structure, which would release
$h\times10$\,GHz and which are relevant for trap loss through
inelastic Li$^*$-Li collisions \cite{kawa93:pra,ritch95:pra}, are
excluded by the repulsive quasi-molecular potential (see Sec.\ 
\ref{subsec:licscol}).

\section{Conclusions}
\label{sec:concl}

Our results can be summarized in the following picture of binary
inelastic Li-Cs collisions in a combined magneto-optical trap. Lithium
and cesium approach each other with a mean relative velocity $\approx
1\,$m/s which is determined by the lithium temperature.  Since the MOT
is operated with near resonant light, atoms can absorb a trapping
photon when the interaction energy is still small compared to $\hbar
\delta$, i.e. at internuclear separations larger than the Condon point
at about 100\,\AA.

When the lithium absorbs a trapping photon at 671\,nm, the excited Li
and ground-state Cs repel each other and inelastic processes are
prevented (optical shielding). The rate coefficient for trap loss by
Li-Cs collisions is therefore found to be independent of the average
Li excitation in the two-species MOT.  When a 852\,nm-photon is
absorbed by the Cs, Li and Cs$^*$ are accelerated by the
attractive molecular potential. Due to the comparatively large
relative velocity, the Cs excitation survives over distances around
300\,\AA. The probability is therefore high to reach very small
internuclear distances in the excited state. The excited
quasi-molecular wavepacket might even oscillate for some periods in
the molecular potential well before spontaneous emission occurs.
Inelastic Li-Cs processes such as changes of the Cs fine-structure
state or the spontaneous emission of a red-detuned photon are then
likely to take place.  Both processes release sufficient energy for
the escape of both atoms from the trap, and our trap loss experiments
do not distinguish among them. The cross section for such inelastic
Li-Cs$^*$ collisions scales with the average Cs excitation in the MOT,
and acquires a value of $\sigma_{\rm CsLi} = \sigma_{\rm LiCs} =
0.7(2)\times10^4$\,\AA$^2$ at maximum excited- state populations
around $\frac{1}{2}$, which corresponds to a trap loss rate
coefficient of $\gamma_{\rm CsLi} = \gamma_{\rm LiCs} =
1.1(2)\times10^{-10}$\,cm$^3$/s.

Collisions involving lithium and cesium in the ground state generally
do not transfer sufficient energy to overcome the trap energy barrier.
If, however, the lithium trap depth is decreased below $\approx
9\,$GHz, lithium atoms eventually escape from the trap after having
undergone a Li-Cs collision in which the cesium changes its hyperfine
ground state. The cesium atom will be retained in the trap since only
5\% of the released energy is transferred to the cesium. Our
measurements yield a cross section larger than $\sigma_{\rm LiCs} =
3\times10^4$\,\AA$^2$ for a ground-state Li-Cs collision with change
of the Cs hyperfine state.  The lower bound for the corresponding rate
coefficient is $\gamma_{\rm LiCs} = 5(1)\times10^{-10}$\,cm$^3$/s.

Homonuclear Cs trap loss collisions give about the same inelastic
cross sections as Li-Cs collisions, while homonuclear Li collisions
have a cross section which is more than one order of magnitude
smaller.  Due to the small extension of the ground-state {\em and} the
excited state interaction potentials, the Li-Cs cross sections are
essentially determined by the s-wave contribution. To our best
knowledge, the short-range part of the Li-Cs molecular potential has
not yet been theoretically investigated. Detailed knowledge on the
fine details of the short-distance molecular potential is necessary to
perform calculations on the relative importance of fine-structure
changing and radiative-escape processes and to estimate trap-loss
cross sections.

Our investigations of inelastic processes between lithium and cesium
represent an important step towards a new class of experiments with
binary atomic mixtures. Such mixtures open new perspectives for
collisional studies in conservative potentials like magnetic or
optical traps, for the formation and trapping of cold polar molecules
through, e.g., photoassociation, or for the investigation of
two-species Bose condensates \cite{law97:prl}.  Starting from our
two-species MOT, we plan to transfer the cold lithium and cesium
simultaneously into a far-detuned optical dipole trap
\cite{grim98:adv} for the investigation of elastic Li-Cs collisions
with the prospect to sympathetically cool lithium with optically
cooled cesium \cite{engl98:applphys}.

\begin{acknowledgement}
  Fruitful discussions with O.~Dulieu are greatfully acknowledged. We
  thank D.~Schwalm for his encouragement and support. This work was
  supported in part by the Deutsche Forschungsgemeinschaft in the
  frame of the Gerhard-Hess-Programm.
\end{acknowledgement}

\end{document}